\documentclass[final,times,3p,twocolumn]{elsarticle}

\usepackage[T1]{fontenc}

\usepackage{lineno,hyperref}
\usepackage{upgreek}
\usepackage{xcolor}
\usepackage{lipsum}
\usepackage{orcidlink}
\usepackage{siunitx}
\usepackage{soul}
\usepackage{subcaption}
\usepackage{bigdelim,bigdelim, makecell, booktabs}
\biboptions{numbers,sort&compress}

\newcommand{\Apr}{A^\prime}

\definecolor{colorOrange}{HTML}{ffa070}
\definecolor{colorRed}{HTML}{c1244f}
\definecolor{colorLightRed}{HTML}{ffcccc}
\definecolor{colorRed}{HTML}{c1244f}
\definecolor{colorLightBlue}{HTML}{add8e6}
\definecolor{colorYellow}{HTML}{ffa500}
\definecolor{colorPink}{HTML}{ff00ff}
\definecolor{colorGreen}{HTML}{33cd32}

\journal{NIMA}

\begin{document}

\begin{frontmatter}

\title{High efficiency veto hadron calorimeter in the NA64 experiment at CERN}
\author[a]{Yu.~M.~Andreev\orcidlink{0000-0002-7397-9665}}
\author[b]{A.~Antonov\orcidlink{0000-0003-1238-5158}}
\author[c,d]{M.~A.~Ayala~Torres\orcidlink{0000-0002-4296-9464}}
\author[e]{D.~Banerjee\orcidlink{0000-0003-0531-1679}}
\author[f]{B.~Banto Oberhauser\orcidlink{0009-0006-4795-1008}\corref{mycorrespondingauthor}}
\cortext[mycorrespondingauthor]{Corresponding authors}
\ead{bantoobb@ethz.ch}
\author[g]{V.~Bautin\orcidlink{0000-0002-5283-6059}}
\author[e]{J.~Bernhard\orcidlink{0000-0001-9256-971X}}
\author[b,h]{P.~Bisio\orcidlink{/0009-0006-8677-7495}}
\author[i]{M.~Bondì\orcidlink{0000-0001-8297-9184}}
\author[b]{A.~Celentano\orcidlink{0000-0002-7104-2983}}
\author[e]{N.~Charitonidis\orcidlink{0000-0001-9506-1022}}
\author[f]{P.~Crivelli\orcidlink{0000-0001-5430-9394}\corref{mycorrespondingauthor}}
\ead{crivelli@phys.ethz.ch}
\author[a]{A.~V.~Dermenev\orcidlink{0000-0001-5619-376X}}
\author[a]{S.~V.~Donskov\orcidlink{0000-0002-3988-7687}}
\author[a]{R.~R.~Dusaev\orcidlink{0000-0002-6147-8038}}
\author[g]{T.~Enik\orcidlink{0000-0002-2761-9730}}
\author[g]{V.~N.~Frolov}
\author[g]{S.~V.~Gertsenberger\orcidlink{0009-0006-1640-9443}}
\author[e]{S.~Girod}
\author[a]{S.~N.~Gninenko\orcidlink{0000-0001-6495-7619}}
\author[l]{M.~H\"osgen}
\author[g]{Y.~Kambar\orcidlink{0009-0000-9185-2353}}
\author[a]{A.~E.~Karneyeu\orcidlink{0000-0001-9983-1004}}
\author[g]{G.~Kekelidze\orcidlink{0000-0002-5393-9199}}
\author[l]{B.~Ketzer\orcidlink{0000-0002-3493-3891}}
\author[a]{D.~V.~Kirpichnikov\orcidlink{0000-0002-7177-077X}}
\author[a]{M.~M.~Kirsanov\orcidlink{0000-0002-8879-6538}}
\author[a,g]{V.~A.~Kramarenko\orcidlink{0000-0002-8625-5586}}
\author[a]{L.~V.~Kravchuk\orcidlink{0000-0001-8631-4200}}
\author[a,g]{N.~V.~Krasnikov\orcidlink{0000-0002-8717-6492}}
\author[c,d]{S.~V.~Kuleshov\orcidlink{0000-0002-3065-326X}}
\author[d]{V.~E.~Lyubovitskij\orcidlink{0000-0001-7467-572X}}
\author[g]{V.~Lysan\orcidlink{0009-0004-1795-1651}}
\author[b,1]{A.~Marini\orcidlink{0000-0002-6778-2161}}
\author[b]{L.~Marsicano\orcidlink{0000-0002-8931-7498}}
\author[g]{V.~A.~Matveev\orcidlink{0000-0002-2745-5908}}
\author[d]{R.~Mena~Fredes}
\author[d,m]{R.~Mena~Yanssen}
\author[n]{L.~Molina Bueno\orcidlink{0000-0001-9720-9764}}
\author[f]{M.~Mongillo\orcidlink{0009-0000-7331-4076}}
\author[g]{D.~V.~Peshekhonov\orcidlink{0009-0008-9018-5884}}
\author[a]{V.~A.~Polyakov\orcidlink{0000-0001-5989-0990}}
\author[o]{B.~Radics\orcidlink{0000-0002-8978-1725}}
\author[g]{K.~Salamatin\orcidlink{0000-0001-6287-8685}}
\author[a]{V.~D.~Samoylenko}
\author[f]{H.~Sieber\orcidlink{0000-0003-1476-4258}}
\author[a]{D.~Shchukin\orcidlink{0009-0007-5508-3615}}
\author[d,p]{O.~Soto}
\author[a]{V.~O.~Tikhomirov\orcidlink{0000-0002-9634-0581}}
\author[a]{I.~Tlisova\orcidlink{0000-0003-1552-2015}}
\author[a]{A.~N.~Toropin\orcidlink{0000-0002-2106-4041}}
\author[n]{M.~Tuzi\orcidlink{0009-0000-6276-1401}}
\author[a,g]{P.~V.~Volkov\orcidlink{0000-0002-7668-3691}}
\author[a]{I.~V.~Voronchikhin\orcidlink{0000-0003-3037-636X}}
\author[c,d]{J.~Zamora-Sa\'a\orcidlink{0000-0002-5030-7516}}
\author[g]{A.~S.~Zhevlakov\orcidlink{0000-0002-7775-5917}}

\affiliation[a]{organization={Authors affiliated with an institute covered by a cooperation agreement with CERN}}
\affiliation[b]{organization={INFN, Sezione di Genova}, postcode={16147}, city={Genova}, country={Italia}}
\affiliation[c]{organization={Center for Theoretical and Experimental Particle Physics, Facultad de Ciencias Exactas, Universidad Andres Bello}, city={Fernandez Concha 700, Santiago}, country={Chile}}
\affiliation[d]{organization={Millennium Institute for Subatomic Physics at High-Energy Frontier (SAPHIR)}, city={Fernandez Concha 700, Santiago}, country={Chile}}
\affiliation[e]{organization={CERN, European Organization for Nuclear Research}, postcode={CH-1211}, city={Geneva}, country={Switzerland}}
\affiliation[f]{organization={ETH Z\"urich, Institute for Particle Physics and Astrophysics}, postcode={CH-8093},city={Z\"urich},country={Switzerland}}
\affiliation[g]{organization={Authors affiliated with an international laboratory covered by a cooperation agreement with CERN}}
\affiliation[h]{organization={Universit\`a degli Studi di Genova}, postcode={16126}, city={Genova}, country={Italia}}
\affiliation[i]{organization={INFN, Sezione di Catania}, postcode={95123},city={Catania}, country={Italia}}
\affiliation[l]{organization={Universit\"{a}t Bonn, Helmholtz-Institut f\"ur Strahlen-und Kernphysik}, postcode={53115}, city={Bonn}, country={Germany}}
\affiliation[m]{organization={Universidad T\'ecnica Federico Santa Mar\'ia and CCTVal}, postcode={2390123}, city={Valpara\'iso}, country={Chile}}
\affiliation[n]{organization={Instituto de Fisica Corpuscular (CSIC/UV)}, city={Carrer del Catedratic Jose Beltran Martinez, 2, 46980 Paterna, Valencia}, country={Spain}}
\affiliation[o]{organization={Department of Physics and Astronomy, York University}, city={Toronto, ON}, country={Canada}}
\affiliation[p]{organization={Departamento de Fisica, Facultad de Ciencias, Universidad de La Serena}, city={Avenida Cisternas 1200, La Serena}, country={Chile}}

\begin{abstract}
NA64 is a fixed-target experiment at the CERN SPS designed to search for Light particle Dark Matter (LDM) candidates with masses in the sub-GeV range. During the 2016-2022 runs, the experiment obtained the world-leading constraints, leaving however part of the well-motivated region of parameter space suggested by benchmark LDM models still unexplored. To further improve sensitivity, as part of the upgrades to the setup of NA64 at the CERN SPS H4 beamline, a prototype veto hadron calorimeter (VHCAL) was installed in the downstream region of the experiment during the 2023 run. The VHCAL, made of Cu-Sc layers, was expected to be an efficient veto against upstream electroproduction of large-angle hadrons or photon-nuclear interactions, reducing the background from secondary particles escaping the detector acceptance. With the collected statistics of $4.4\times10^{11}$ electrons on target (EOT), we demonstrate the effectiveness of this approach by rejecting this background by more than an order of magnitude. This result provides an essential input for designing a full-scale optimized VHCAL to continue running background-free during LHC Run 4, when we expect to collect $10^{13}$ EOT. Furthermore, this technique combined with improvements in the analysis enables us to decrease our missing energy threshold from $\SI{50}{GeV}$ to $\SI{40}{GeV}$ thereby enhancing the signal sensitivity of NA64.
\end{abstract}

\begin{keyword}
Light Dark Matter \sep Fixed-Target Experiment \sep Missing-Energy Experiment \sep Hadronic Calorimeter
\end{keyword}

\end{frontmatter}

\section{Introduction: the NA64 experiment at CERN}\label{sec:Introduction}

In the quest for understanding the nature of Dark Matter (DM), thermal relic models stand as one of the most appealing theoretical frameworks. In the case of freeze-out models, dark and ordinary matter are assumed to be in thermal equilibrium in the early Universe, and, as the Universe expands, the DM mass density approaches the observed DM abundance \cite{Kolb:1990vq,Rubakov:2017xzr,Madhavacheril:2013cna}. The DM relic density imposes stringent constraints on the masses and interactions strengths with SM particles, leading to highly-predictive models \cite{Feng:2008ya}. If, additionally, DM is embedded within a new Dark Sector (DS) that introduces a new force mediator, masses in the MeV$-$GeV range become possible \cite{Batell:2009di,Pospelov:2008zw,Essig:2013lka,Batell:2014mga}. Therefore, the thermal history of these scenarios sets explicit regions of interests that serve as clear targets for experiments searching for Light DM (LDM) in the sub-GeV mass range \cite{Battaglieri:2017aum,Akesson:2018vlm,Lanfranchi:2020crw}.

The NA64 experiment has been operating since 2016 in the H4 beamline at the CERN North Area facility \cite{Atherton:164934,Banerjee:2774716}, exploiting the high-purity electron beam in the search for LDM below the electroweak scale \cite{Andreas:2013lya,Banerjee:2017hhz,Andreev:2023uwc}. NA64 is a fixed-target experiment, searching for rare events where the production of a DS mediator takes place in the active beam dump. The DS particle produced this way would decay into a pair of LDM particles, carrying away some of the initial electron's energy and leaving the setup undetected. Therefore, the signature in the invisible searches at NA64 is characterized by missing energy from a clean electron primary impinging on the ECAL.

\begin{figure}[t!]
    \centering
    \includegraphics[width=.45\textwidth]{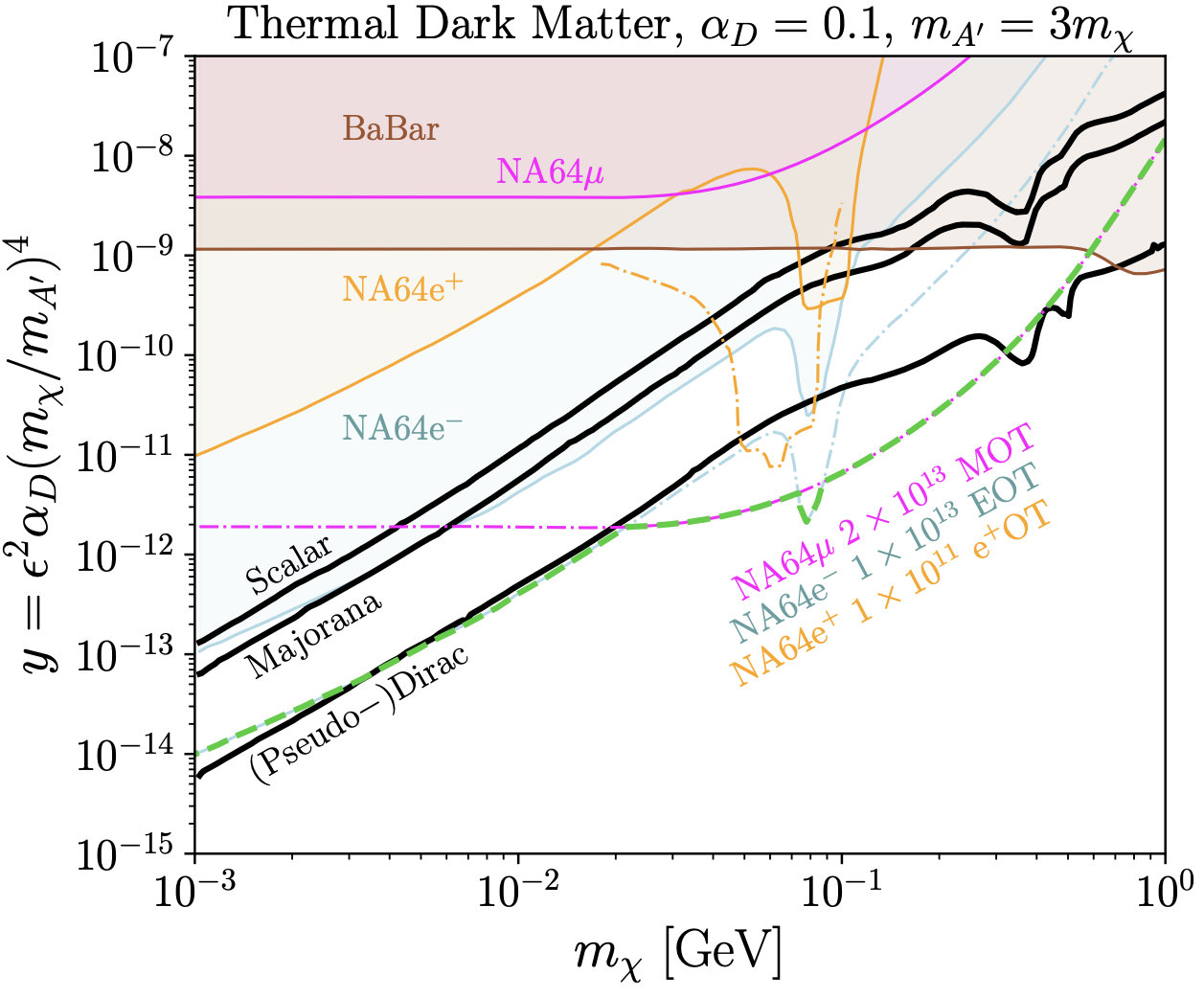}
    \caption{Current NA64 $90\%$ C.L. exclusion limits on the Dark Photon scenario $\Apr\rightarrow$~invisible, in the exploration with $e^{-}$ (solid \textcolor{colorLightBlue}{light blue} line), $e^{+}$ (solid \textcolor{colorYellow}{yellow} line) and $\mu$ (solid \textcolor{colorPink}{pink} line) beams \cite{NA64:2024nwj}. The combined projection is displayed as a \textcolor{colorGreen}{green} dashed line and is obtained from the projections of the individual programs (dash-dotted lines with the respective colors) for the statistics to be collected during the LHC Run 4.}
    \label{fig:ldm}
\end{figure}

Having found no trace of any signal-like events in $9.37\times10^{11}$ electrons on target (EOT), NA64 has been able to place the current most binding exclusion limits in case of an invisibly-decaying new vector boson for the mass range $m_{\chi}\approx\SI{1}{MeV}-\SI{100}{MeV}$ \cite{Andreev:2023uwc}, as illustrated in Fig.~\ref{fig:ldm}. In the forthcoming years, NA64 aims to accumulate about $10^{13}$ EOT to decisively probe the LDM parameters motivated by the thermal targets \cite{Crivelli:2907892}. In light of this, a detector upgrade is planned during the next CERN long shutdown (LS3).
 
\section{NA64 approach to LDM and experimental setup}\label{sec:NA64}

In the most representative models, the new force mediator is a vector boson $\Apr$ called the dark photon. If $\Apr$, exists, it would be produced in high-energy $e^{-}$ scattering with the target nuclei via the reaction $e^{-} \mathcal{N} \rightarrow e^{-} \mathcal{N} \Apr$ or through resonant annihilation of secondary positrons from the electromagnetic (e-m) shower $e^{-}e^{+}\rightarrow\Apr$. The subsequent decay $\Apr\rightarrow\mathrm{LDM}$ would lead to an event with large missing energy as the LDM particles escape detection.

\begin{figure*}[t!]
    \centering
    \includegraphics[width=.95\textwidth]{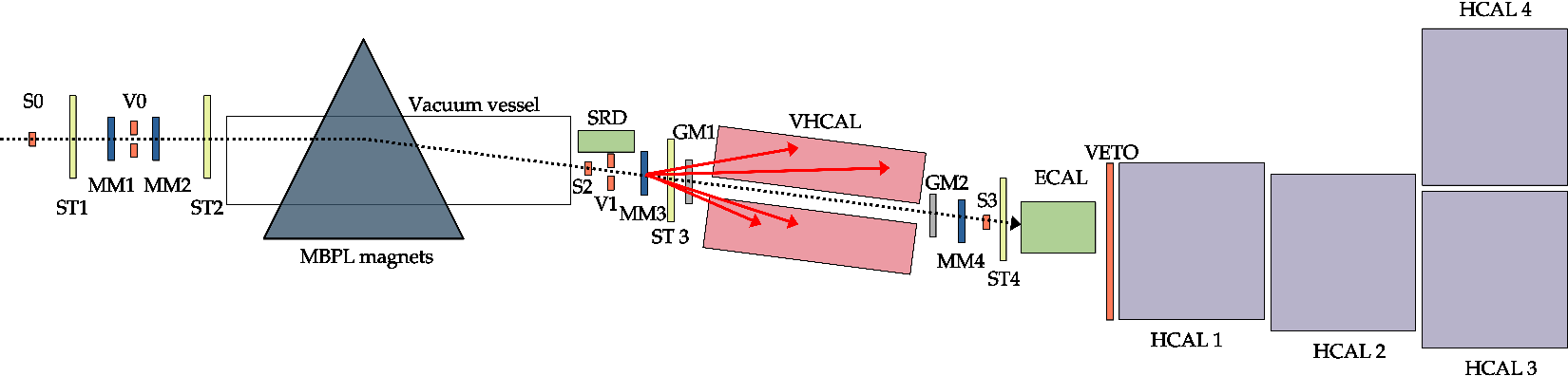}
    \caption{The NA64 setup in the invisible mode configuration during the 2023 run at the H4 beamline. The red arrows correspond to a schematic representation of the particle shower as a result of upstream electron-nuclear interactions in one of the Micromegas trackers by the primary electron. See text for further details.}
    \label{fig:detector}
\end{figure*}

The LDM searches at NA64 therefore rely on the detection of the impinging beam electron on the target, requiring a clear definition of its incoming track and the effective containment of the interactions inside the target. For this purpose, the NA64 detector, as shown in Fig.~\ref{fig:detector} consists of the following sub-systems \cite{Andreev:2023uwc}: (I) a series of scintillator counters (Sc) that define the beam and trigger ($S_{0}$, $S_{2}$, $S_{3}$, $V_{0}$, $V_{1}$), (II) a magnetic spectrometer to reconstruct the incoming momenta of the $\SI{100}{GeV}$ beam electrons, with Straw chambers \cite{Volkov:2019qhb} (ST), Gas Electron Multiplier (GEM) and Micromegas (MM) detectors \cite{Banerjee:2017mdu} (III), a synchrotron radiation detector (SRD) to tag incoming electrons based on their emission of SR photons as their trajectory is bended by the magnetic field \cite{Depero:2017mrr}, (IV) a $19\times23\times\SI{47}{cm^3}$, $40$ radiation lengths ($X_{0}$) Pb/Sc electromagnetic calorimeter (ECAL) that serves as the active target, (V) one large high-efficiency scintillator counter (VETO) that vetoes charged particles produced mainly in the ECAL, and (VI) four $60\times60\times\SI{163}{cm^3}$ hadronic calorimeters (HCAL) that, given the large Lorentz boost of the incoming electrons, provide the necessary hermeticity to ensure the detection of secondary hadrons produced in the target. The physics data is collected using the trigger signal $\mathcal{S}_{trig}$ from the beam-defining trigger scintillators (Sc) from point (I), with two additional ECAL: a minimum energy deposit of $\mathcal{S}_{PRS}:=\left(E_{\mathrm{PRS}}\gtrsim\SI{300}{MeV}\right)$ in the first $4X_{0}$ that serve as the pre-shower (PRS) and a maximum energy deposit of $\mathcal{S}_{EC}:=\left(E_{\mathrm{EC}}\lesssim\SI{80}{GeV}\right)$ in the main ECAL. We therefore define the production trigger as $\mathcal{S}_{phys} := \mathcal{S}_{trig}\times\mathcal{S}_{PRS}\times\mathcal{S}_{EC}$.

\begin{figure}[ht!]
    \centering
    \includegraphics[trim={1px 1px 1px 1px},clip,width=0.45\textwidth]{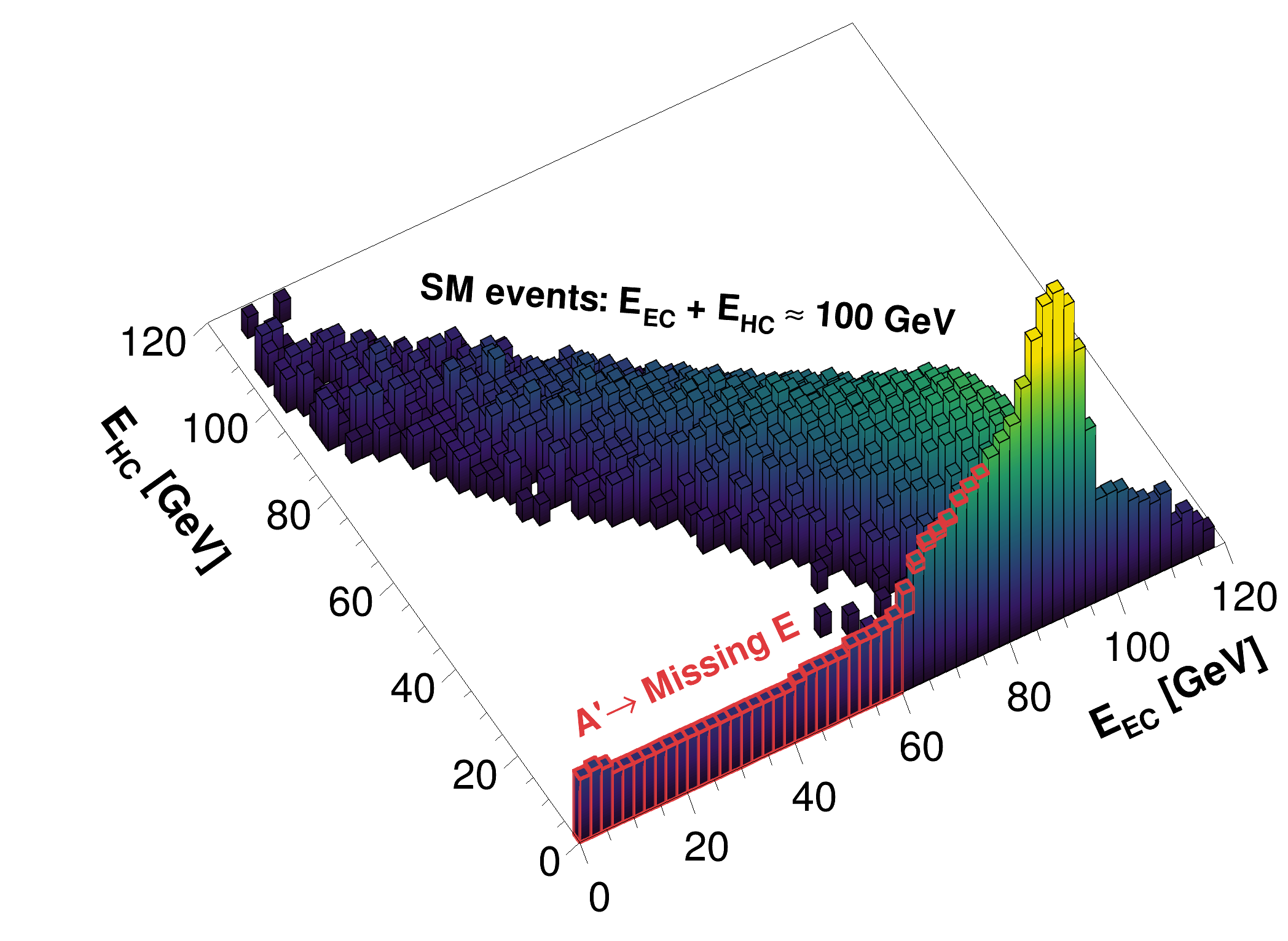}
    \caption{Example of the distribution of physics data events in the $\left(E_{\mathrm{EC}}; E_{\mathrm{HC}}\right)$ plane, including simulated events with $\Apr$ production in the target (highlighted in \textcolor{colorRed}{red}). Events in the diagonal correspond to SM processes.}
    \label{fig:herm_lego}
\end{figure}

The detector response in the case of $\Apr$ production is illustrated in Fig.~\ref{fig:herm_lego}, showing the distribution of signal events in the $\left(E_{\mathrm{EC}}; E_{\mathrm{HC}}\right)$ plane, obtained by simulating the interaction of $\SI{100}{GeV}$ electrons impacting on the ECAL using the DGM4 package \cite{Bondi:2021nfp, Oberhauser:2024ozf}. Signal events, highlighted in red, would leave significant missing energy in the active target, $E_{ECAL}$, and no trace of their passage through the HCAL, $E_{HCAL}$. For SM interactions, the hadron calorimeters would detect particles leaking target, thus satisfying energy conservation $E_{\mathrm{EC}}+E_{\mathrm{HC}}\approx\SI{100}{GeV}$.

The NA64 detector is designed to have a high background rejection while maintaining an optimal signal efficiency. To minimize the background from SM events, a candidate event with the production of an invisible DS mediator needs to pass the following criteria: (i) The detected synchrotron radiation (SR) should be detected in all three modules of the SRD in time with the trigger and be greater than $\SI{1}{MeV}$, (ii) the incoming track should have a reconstructed momentum in the range $\SI{100 \pm 6}{GeV}$, (iii) the track's entrance and exit angles from the magnetic spectrometer should be within $\SI{1}{mrad}$ and the extrapolation of the track should reach the central cell of the target, (iv) the energy deposition in the ECAL, including the longitudinal and transverse shape of the shower, should be consistent with an impinging electron, (v) no activity should be recorded in the zero-degree hadron calorimeter (HCAL$_{4}$), (vi) no energy deposition is detected in the VETO, (vii) there should be only one clean hit in the Straw chambers downstream, ST3 and ST4, and (viii) no energy deposition is observed in the hadron calorimeters, VHCAL and HCAL. In particular, the hadron contamination present in the beam at H4 has been extensively studied, with the admixture $h/e^{-}$ estimated to be at the level $\lesssim0.3\%$ \cite{Andreev:2023xmj}. The hadronic background is then further suppressed by particle identification based on the SR radiation \cite{Depero:2017mrr}. Finally, the low level of track mis-reconstruction allows an accurate determination of the incoming momentum and effective rejection of low energy $e^{-}$ remnants \cite{Banerjee:2017mdu}.

The main motivation for this work is provided by the most recent publication \cite{Andreev:2023uwc}, in which hadrons from electroproduction on the upstream beamline material (EUM) and escaping the HCAL acceptance were estimated to be one of the most significant contributions to the total expected background, limiting further improvements in the search sensitivity. For this reason, in 2023 and 2024 a prototype veto hadron calorimeter with a hole was placed in the beamline before the target and used to reject events with large-angle scattering hadrons that escape the detection by the ECAL and HCAL. Preliminary results for the analysis of the 2023 run indicate that the prototype VHCAL is successful and ostensibly reduces the background from escaping hadrons.

In this work, we report on the detailed study of the performance of the prototype VHCAL installed in order to address the background from hadrons produced via electron-nuclear and photon-nuclear interactions. Specifically, we share results from the analysis of the 2023 run and validate with these our Monte Carlo (MC) framework. Then, different configurations in MC are studied, necessary to drive the upgrades considered for the next years. Finally, we include results using a setup with a full-scale VHCAL module, delineating the first steps towards an optimized VHCAL module.

\section{Background from hadrons via upstream electroproduction} \label{sec:Hadrons ElectronNuclear}

The primary beam electrons traversing the setup can interact with the material placed upstream of the ECAL and through deep inelastic scattering with the nuclei produce a shower of hadrons, $e^{-} \mathcal{N} \rightarrow e^{-} \mathcal{N}^{\prime} X$, where $\mathcal{N}$ is the nucleus in a given material and $X$ corresponds to the hadronic final state. At $\SI{100}{GeV}$, beam electrons can produce numerous hadrons through deep inelastic scattering on nuclei. Most of the produced secondary particles are pions, followed by heavier hadrons such as $K$, $\eta$, $\eta^{\prime}$, $p$ and $n$. These hadrons may also be produced upstream of the target via $\gamma$-nuclear processes. This is the case if a photon is emitted by the beam electron through Bremsstrahlung and then proceeds to interact with the nucleus in the material present upstream.

The probability of such interactions has been minimized by reducing the material budget along the beamline as well as placing a $\SI{16}{m}$-long vacuum vessel. In particular, the total material budget between the end of the vacuum vessel and the ECAL is estimated to be $0.06$ radiation lengths ($X_{0}$), with the largest contributions originating from the trackers GM1, GM2, MM3 and MM4. In any case, the emission angle of these secondary hadrons may be large enough to escape any detection in the hadronic calorimeters. This lack of hermeticity is enhanced if we consider a neutral hadron, such as $K^{0}_{L}$, which also evades the detection of our trackers and VETO. Events with these characteristics may mimic the potential missing-energy signal and contribute to the expected background of the invisible searches at NA64.

While $e^{-}$-nuclear and $\gamma$-nuclear interactions can take place in the many elements upstream of the ECAL, as sketched in Fig.~\ref{fig:detector}, not all of these interactions contribute in the same way to our background. The background from hadrons produced before the magnetic spectrometer is heavily suppressed, as the scattered electron needs to have sufficient momentum to traverse the trigger scintillators $S_{2}$ and $S_{3}$ downstream after the deflection from the dipole magnets. Since the electron's emitted synchrotron radiation scales with its kinetic energy to the fourth power, the cut on the SRD is extremely effective at suppressing these events \cite{Depero:2017mrr}. Additionally, any neutral hadrons or photons produced this way need to miss the SRD and the zero-degree hadron calorimeter to be categorized as dangerous candidate events. Furthermore, the background from $\gamma$-nuclear interactions due to a high-energy Bremsstrahlung photon is expected to be comparable\footnote{The cited work from Y. Tsai indicates that for an electron beam the production of particles via a real Bremsstrahlung photon is comparable to the contribution from virtual photons from electroproduction for a target of thickness $t_{eq} \approx 1/50\, X_{0}$.} to the electroproduction given a thickness of $0.06 \,X_{0}$ \cite{Tsai:1973py}. Therefore, we expect the background from escaping particles to be dominated by hadrons produced via $e^{-}$-nuclear and $\gamma$-nuclear interactions between the end of the vacuum vessel and the start of the ECAL.

\begin{figure*}[t]
\begin{center}
\includegraphics[width=0.97\textwidth]{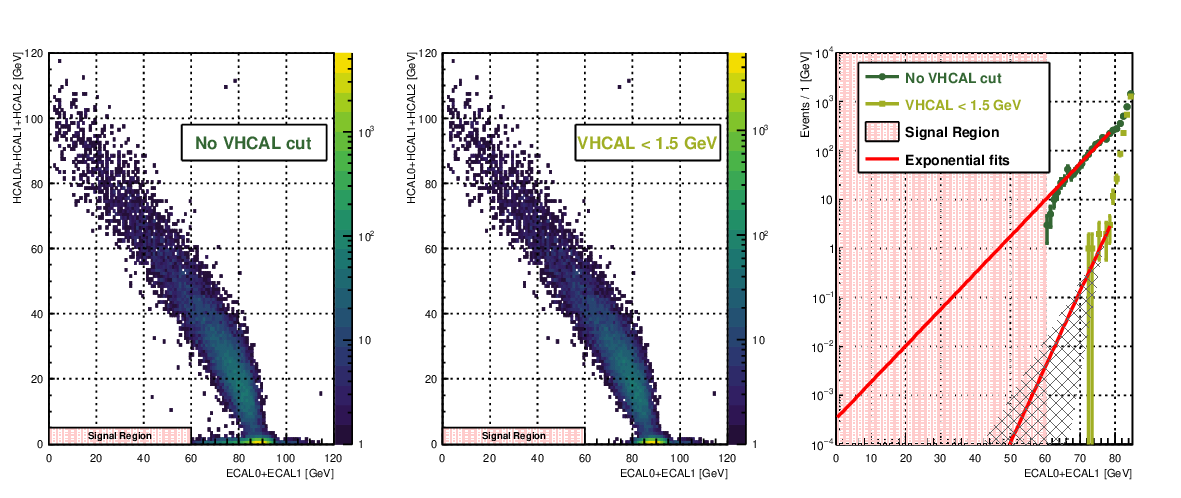}
    \caption{Distribution of events in the $\left(E_{\mathrm{EC}}; E_{\mathrm{HC}}\right)$ plane before (left) and after (middle) applying a cut on the total VHCAL energy for a subset of the total statistics collected during the 2023 run, with the signal-region blinding condition still applied, after all other selection criteria are applied. The impact of the cut in the estimated number of background events is highlighted by the extrapolation of the tails in the sideband $E_{\mathrm{HC}}<\SI{1.5}{GeV}$ (right). Most of the events removed by the VHCAL cut lie close to the signal box and likely correspond to secondary hadrons that escape the HCAL acceptance. The gray crosshatched region represents the systematic uncertainty from the fitting procedure, defined by the area between fits with the minimum and maximum integral obtained by varying the fitting range.}
    \label{fig:hermeticity_VHCAL_diff} 
\end{center}
\end{figure*}

After applying the selection (i)-(vii) as defined in Section~\ref{sec:NA64}, we can identify events with secondary hadrons produced upstream of the ECAL by their energy deposition in the calorimeters. In Figure~\ref{fig:hermeticity_VHCAL_diff}, the bi-dimensional distribution of the HCAL and ECAL energies is shown for a subset of the physics data collected in 2023. Events in the bi-plots are mostly composed of $e^{-}$-nuclear and $\gamma$-nuclear interactions before or inside the ECAL, as well as the low-energy tail from $\SI{100}{GeV}$ electrons due to the ECAL resolution. EUM events with secondary hadrons are likely to pass the online production trigger condition if the scattered electron is caught by the ECAL and the rest of the energy is either lost or deposited in the hadronic calorimeters. Given that the VETO cut is applied, events shown in the bi-plot with significant HCAL energies are more likely to be secondary neutrals that are detected by the first HCAL module, see Fig.~\ref{fig:detector}. Hadrons created inside the target via electron- or photoproduction in the lead converter can leak out of the ECAL, as the equivalent nuclear interaction length of this volume is $\lambda \approx 1.56\lambda_{0}$. These hadrons are reliably detected by the HCAL modules placed after, and so the total energy is conserved, which corresponds to the band of events in $\left(E_{\mathrm{EC}}+E_{\mathrm{HC}}\right)\approx\SI{100}{GeV}$.

In addition to the energy deposition in the ECAL and HCAL, low-energy charged hadrons in the showers from upstream $e^{-}$-nuclear and $\gamma$-nuclear interactions can be detected by the trackers. In this case, interactions upstream of the target can yield an increased multiplicity of hits in the trackers closest to the production vertex. The Straw chambers, which have the largest acceptance, are particularly sensitive to this and for this purpose, in the invisible analysis selection, we apply a cut on the multiplicity of hits in ST3 and ST4 to improve the rejection of charged secondaries produced upstream of the ECAL. On the other hand, neutral hadrons evade completely this detection, and therefore this method can only complement the hermeticity provided by the VHCAL and the HCAL.

Finally, the EUM contribution to the expected background can be derived using the remaining events with no HCAL energy in the sideband $E_{\mathrm{HC}}<\SI{1.5}{GeV}$, as illustrated in the rightmost plot in Fig.~\ref{fig:hermeticity_VHCAL_diff}. Events in the sideband are fit with an exponential PDF, and the latter is integrated within the signal region to obtain a data-driven estimate of the expected yield. The stability of this estimate and the corresponding uncertainty are assessed by repeating the procedure varying the fit range. As expected, applying a cut on the VHCAL energy predicts a significantly low number of background events in 2023: $n_{b}=\SI[parse-numbers=false]{(1.1 \pm 1.4 (stat) \pm 2.4 (sys))\times10^{-2}}{}$ for $1.232\times10^{11}$~EOT. However, this estimate carries a relatively large systematic uncertainty associated with the fit procedure, as shown by the gray crosshatched region in Fig.~\ref{fig:hermeticity_VHCAL_diff}. Therefore, an alternative approach is required to reliably optimize the detector and to obtain a robust projection for the expected level of this background for $10^{13}$~EOT.

\section{Overview of 2023 run}\label{sec:Overview 2023}

During the data taking in 2023, the prototype VHCAL was installed just before the main target as shown in Fig. \ref{fig:detector}. The VHCAL was slightly rotated in order to place it parallel to the bent beam. The rotation angle was estimated to be around $\SI{32}{mrad}$ with respect to the beam axis from measurements of the 2023 setup provided by the Geodetic Metrology group (BE-GM) at CERN. Its position minimizes the distance between the material in the downstream region of the experiment and the first hadronic calorimeter, reducing the probability of escaping hadrons. The distance between the vacuum window and the closest hadronic calorimeter (HCAL in 2022 and VHCAL in 2023) was reduced from $\SI{3.845}{m}$ to $\SI{0.784}{m}$. In 2023, the largest distance between a detector and a hadronic calorimeter is the one between the second GEM tracker (GM2) and the first HCAL module, approximately $\SI{1.711}{m}$. The size of the central hole is large enough to avoid interactions with the beam, considering that the beam spot is defined by the $\varnothing\SI{32}{mm}$ trigger scintillators. The aperture is also sufficiently narrow that particles passing through it are covered by the acceptance of the HCAL.

\subsection{Prototype VHCAL}\label{subsec:Experimental Setup}

\begin{figure}[ht!]
\centering
\begin{subfigure}{.235\textwidth}
\centering
\includegraphics[trim={0 2.02cm 0 0}, clip, width=0.95\textwidth]{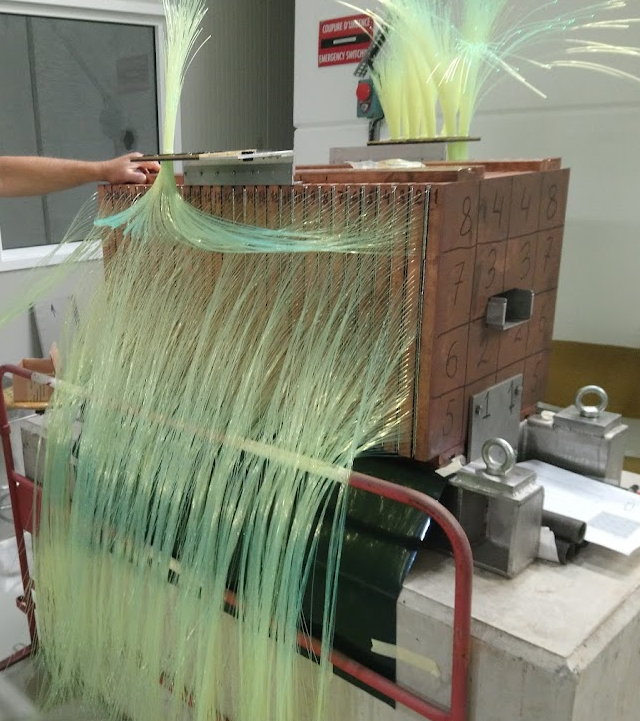}
\end{subfigure}%
\begin{subfigure}{.235\textwidth}
\centering
\includegraphics[trim={0 3cm 0 1.38cm}, clip, width=0.95\textwidth]{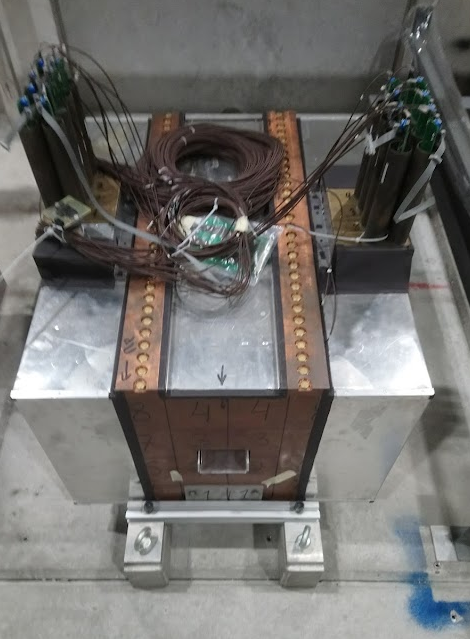}
\end{subfigure}
    \caption{Images of the copper prototype VHCAL before (left) and after (right) its assembly in 2021.}
    \label{fig:vhcal_picture} 
\end{figure}

The prototype VHCAL is a $500\times500\times\SI{1000}{mm^3}$ copper hadronic calorimeter that has a central hole with dimensions $120\times\SI{60}{mm^2}$. It consists of 30 layers of $\SI{25}{mm}$ copper plates and 29 layers of $\SI{2}{mm}$ scintillating material. It is divided into an array of $4\times4$ cells and, in terms of its material budget, its longitudinal depth corresponds to approximately $\lambda\approx5$ nuclear interaction lengths. In the same manner as it's the case for the ECAL and HCAL modules, each cell is connected through wavelength-shifting (WLS) fibers to a photo-multiplier tube (PMT) readout. The number of photo-electrons (phe) was measured to be approximately $20$ in a single \SI{2}{mm} scintillator layer for a minimum ionizing particle (MIP) signal from cosmic rays, corresponding to a yield of $\SI{50}{\mathrm{phe}\per MeV}$. Therefore, the total for a MIP passing through a VHCAL cell is approximately $\SI{600}{\mathrm{phe}}$. The readout of each PMT is performed by two interleaved sampling ADCs (MSADC) \cite{Mann:2009ieee,Huber:2011zza} including a shaper. The sampling frequency obtained from interleaving the two ADCs is $\SI{80}{MHz}$, corresponding to $\SI{12.5}{ns}$ between samples, with a 12 bit resolution for the amplitudes. Pictures of the VHCAL before and after assembly can be seen in Fig.~\ref{fig:vhcal_picture}. 

At the beginning of the 2023 run, each VHCAL cell was simultaneously calibrated using a wide $\SI{100}{GeV}$ $\mu$ beam and adding the VETO to the trigger logic. With the VETO in the trigger, the detected beam of muons effectively covers the whole VHCAL module. The energy deposition in each cell is then compared to that expected from a MIP ($E_{\mathrm{MIP}}\approx\SI{1.5}{GeV}$) to obtain the calibration factors. The same calibration data is used to determine the expected timing for each VHCAL cell with respect to the trigger. Given that the resulting resolution is of the order of $\sigma_{t}\approx\SI{2}{ns}$, we apply a time window of $|t-t_{0}| < 5\sigma_{t} \approx \SI{10}{ns}$ when selecting in-time energy depositions.

To correct these calibration factors resulting from the variations in the PMT gain, the response in each cell was monitored throughout the 2023 run. This is done by continuously measuring the PMT's response to a LED light between spills. We include this spill-by-spill correction factor in the calculation of the energy from the maximum amplitude of the waveform.

\subsection{Data analysis for 2023 run}\label{subsec:Data analysis}

Prior to its introduction to the setup, the prototype VHCAL was expected to improve the hermeticity of the detector. As the background contribution from escaping hadrons created by interactions before the target is estimated from data, the extent of this improvement can only be obtained by looking at the data from the 2023 run. Preliminary results for a sub-period corresponding to $1.232\times10^{11}$~EOT are displayed in the middle plot of Fig.~\ref{fig:hermeticity_VHCAL_diff}. As the analysis on invisible signatures is still ongoing, events in the signal region corresponding to $E_{\mathrm{EC}}<\SI{60}{GeV}\land E_{\mathrm{HC}}<\SI{5}{GeV}$ have been blinded. The plot on the right after applying all cuts demonstrates the role of this calorimeter, as a cut on the total VHCAL energy completely removes all events below the diagonal defined by $\left(E_{\mathrm{EC}}+E_{\mathrm{HC}}\right)<\SI{70}{GeV}$.

Current estimates from this analysis indicate a reduction of the EUM background by at least an order of magnitude. Critically, a data-driven estimate of the expected background for $10^{13}$~EOT is constrained by the limited number of remaining events, leading to a significant uncertainty in extrapolating to the signal region. This emphasizes the importance of studying this background through simulations.

\begin{figure}[ht!]
\begin{center}
\includegraphics[width=0.475\textwidth]{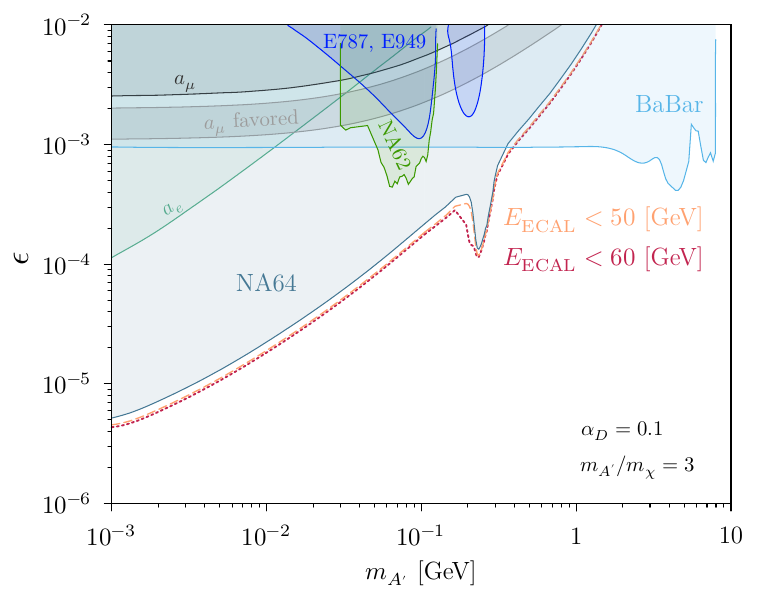}
    \caption{Preliminary NA64 $90\%$ C.L. exclusion limits (dashed lines) obtained in case of no observed signal for the dark photon ($\Apr$) to invisible search in the $(m_{\Apr}, \epsilon)$ plane for the combined 2016-2023. The exclusion limits are drawn for two choices of the signal region in the ECAL: $E_{\mathrm{EC}}<\SI{50}{GeV}$ in \textcolor{colorOrange}{orange} and $E_{\mathrm{EC}}<\SI{60}{GeV}$ in \textcolor{colorRed}{red}. This is only applied to the 2023 dataset and showcases the improved sensitivity enabled by the prototype VHCAL.}
    \label{fig:exclusion_limits_2023} 
\end{center}
\end{figure}

In any case, this major reduction in the overall expected background has allowed extending the signal region from $E_{\mathrm{EC}}<\SI{50}{GeV}$ to $E_{\mathrm{EC}}<\SI{60}{GeV}$. This effectively translates to a higher signal sensitivity in the invisible searches, which is further enhanced for the resonant annihilation production of a dark photon \cite{Andreev:2021fzd}. This is because NA64's sensitivity to the resonant production is directly constrained by the missing energy threshold in the ECAL $E^{miss}_{\mathrm{EC}}$ \cite{Andreev:2021fzd}:

\begin{equation}
    \sqrt{2m_{e}E^{miss}_{\mathrm{EC}}} \lesssim m_{\Apr} \lesssim \sqrt{2m_{e}E_{0}}
\end{equation}

where $m_{e}$ is the electron's mass, $m_{\Apr}$ the dark photon's mass and $E_{0}$ the beam energy. To illustrate this point, the preliminary projections for the $90\%$ C.L. exclusion limits assuming no signal and zero background are depicted for two definitions of the signal region, $E_{\mathrm{EC}}<\SI{50}{GeV}$ and $E_{\mathrm{EC}}<\SI{60}{GeV}$, in Fig.~\ref{fig:exclusion_limits_2023}. Extending the signal region effectively allows the detection of lower mass mediators produced in this resonant process, down to $m_{\Apr}\approx\SI{200}{MeV}$. 

\section{VHCAL study description}\label{sec:Analysis description}

To properly understand the impact of the VHCAL and study a possible optimized configuration, we proceed as follows. First, we conduct a detailed MC simulation of electron-nuclear interactions in the 2023 geometry and compare the resulting sample of events with the data. Specifically, we compare the remaining events after the invisible selection criteria described in Section~\ref{sec:NA64} as well as for a selection of EUM events with charged hadrons. For this, we use ROOT, \texttt{RooFit} and ROOT's \texttt{RDataframe} \cite{Brun:1997pa}\footnote{ROOT version 6.30/02.}. Then, having validated our simulation, we study three different VHCAL configurations and evaluate them in terms of their hermeticity and suppression for this particular background. Finally, we conclude by estimating the EUM background $n_{b}$ for the expected statistics to be collected in the next years. 

\begin{figure}[ht!]
\begin{center}
\includegraphics[trim={1px 0px 0px 1px},clip,width=0.475\textwidth]{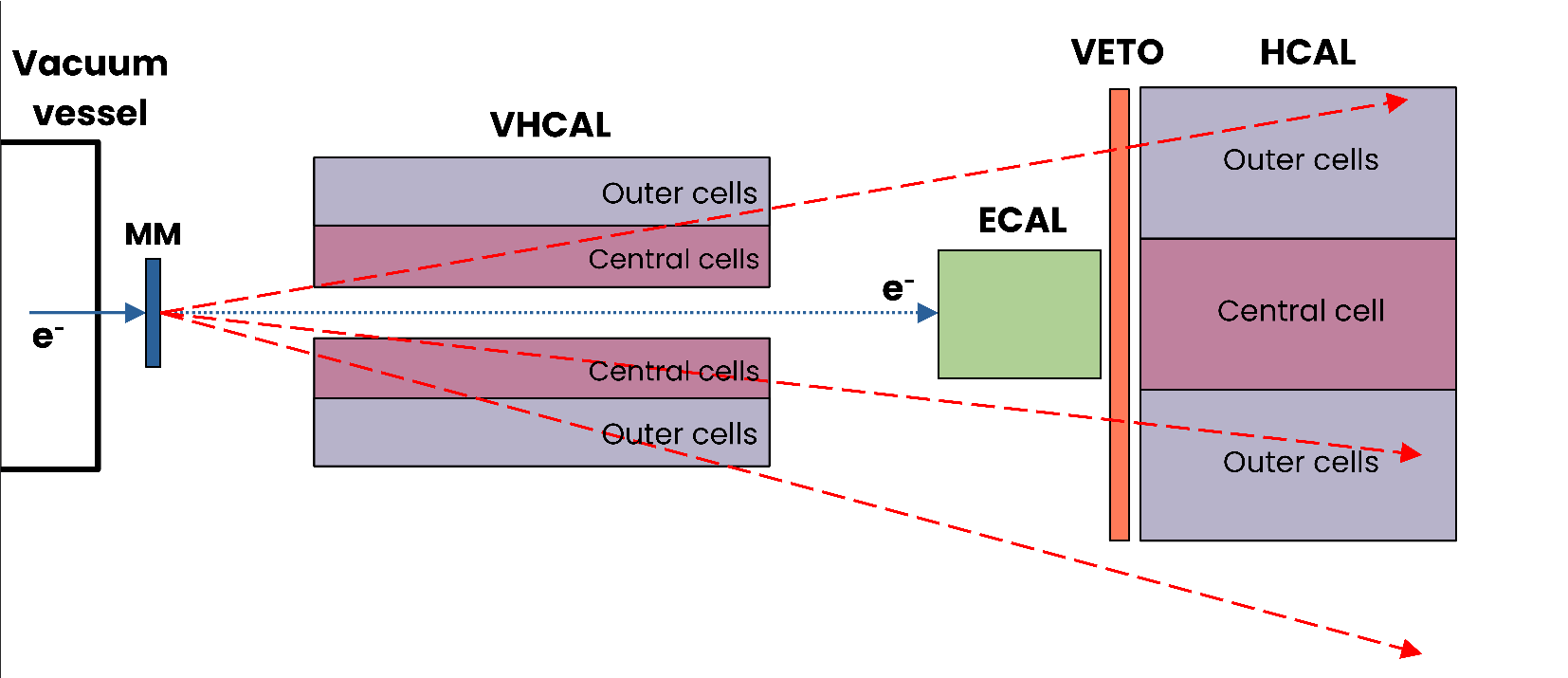}
    \caption{Sketch of an event with electroproduction of hadrons in the MM tracker placed between the vacuum vessel and the VHCAL. Drawn in dashed red arrows are the trajectories of possible hadrons with high transverse momentum. The blue, dotted arrow follows the scattered electron that impinges on the ECAL.}
    \label{fig:VHCAL_coverage} 
\end{center}
\end{figure}

Events with secondary hadrons produced upstream of the prototype VHCAL as the one sketched in Fig.~\ref{fig:VHCAL_coverage} are characterized by hadronic showers in both VHCAL and HCAL. In particular, the energy deposition is expected to be mostly in the central cells of the VHCAL, while in the first HCAL module the opposite is the case. In order to assess how well the MC reproduces the emission angle of these hadrons, we define two quantities that reflect the spatial topology of the energy deposition in the hadron calorimeters. For the VHCAL we define the fraction of energy deposited in the twelve outermost cells as $R_{\mathrm{VHC}}$:

\begin{equation}
    R_{\mathrm{VHC}} = \frac{E_{\mathrm{VHC}} - E_{\mathrm{VHC}, 2\times2}}{E_{\mathrm{VHC}}}
    \label{eq:rvalue_vhcal}
\end{equation}

Similarly, we establish this ratio in an HCAL module by 

\begin{equation}
    R_{\mathrm{HC}} = \frac{E_{\mathrm{HC}} - E_{\mathrm{HC, center}}}{E_{\mathrm{HC}}}
    \label{eq:rvalue_hcal}
\end{equation}

where $E_{\mathrm{HC, center}}$ is the energy deposition in the central HCAL cell.

As both of these ratios are sensitive to contributions from noise, an energy threshold cut is applied to each calorimeter cell when computing this value. This means that energy depositions below the observed level of noise in data, $E_{\mathrm{Cell}}<\SI{300}{MeV}$, are set to $0$.

\subsection{Monte Carlo simulations}\label{subsec:MC}

Having observed the effectiveness of the prototype VHCAL in 2023, we developed MC simulations to study the potential of a full-scale VHCAL module to be assembled during LS3. For this purpose, samples of events were generated using the Geant4-based~\cite{Agostinelli:2002hh,Allison:2006ve, ALLISON2016186} simulation framework of NA64\footnote{Geant4 version: 4.11.02} to study the characteristics of the hadrons produced via electron-nuclear and photon-nuclear interactions. The MC samples were obtained from simulations with a realistic geometry and beam definition based on the setup configuration of 2023, using the positions of all the elements in the H4 beamline setup as measured by the BE-GM at CERN. Particular care was taken to correctly implement the structure of the different trackers and Sc counters, which are the main source of photon-nuclear and electron-nuclear interactions upstream of the ECAL. Additionally, the simulations incorporated a detailed model of the dipole magnets' magnetic fields, while the initial beam momentum and position distributions were derived from calibration trigger events by extrapolating the fitted tracks back to the start of the setup.

To reach a statistic larger than the $4.38\times10^{11}$~EOT collected in 2023, a bias factor is introduced to the production cross-section for $e^{-}$-nuclear and $\gamma$-nuclear interactions using the Generic Event Biasing in Geant4 \cite{G4bias}. In particular, we attach a bias factor for these interactions to the logical volumes of the setup elements that are located upstream with respect to the ECAL. To prevent chains of biased interactions, the bias applied to the selected processes is exclusively restricted to \texttt{G4track} instances with weights that are not the result of a biased interaction.

Finally, to provide a valid comparison, MC events are processed using the same event reconstruction pipeline as for data. The reconstructed hits in the tracking detectors are then fitted to physical tracks with the \texttt{GenFit} package \cite{Rauch:2014wta} using a deterministic annealing filter. Additionally, we apply a fuzzy logic filter to the MC sample, in order to replicate the effect of the online production trigger condition $\mathcal{S}_{phys}$. The ECAL energy selections, $\mathcal{S}_{PRS}$ and $\mathcal{S}_{EC}$, are applied on the uncalibrated PMT response, meaning they do not correspond directly to a cut on the calibrated energy deposition in PRS and ECAL, respectively. Instead, we model this effect by fitting a sigmoid function to the observed ratio of events passing the online selection $\mathcal{S}_{phys}$ to those passing the calibration trigger selection $\mathcal{S}_{trig}$, as a function of PRS or ECAL energy. Then, for each MC event, the energy-dependent value from the sigmoid is compared to a uniformly distributed random number in the interval $[0,1]$. If the function value is greater, the event is kept; otherwise, it is discarded.

\subsection{Monte Carlo validation}\label{subsec:MC validation}

To address the quality of the simulation, we benchmark the detector response in MC against data from the 2023 run by selecting in both cases EUM events. The selection criteria are applied with the same cut values, with a few exceptions. Namely, the cut based on the e-m shower in the ECAL is not applied to either MC or data as part of the selection (iv). This is because the expected e-m shower shape is determined using calibration events, and the discrimination of the observed e-m shower is purely data-driven. A proper implementation of this cut in MC requires a rigorous validation of the e-m showers in the ECAL that is beyond the scope of this work. Furthermore, we evaluate the impact of the initial beam definition on the resulting efficiencies for MC and data by estimating the systematic uncertainty associated with the difference in the efficiency of the selection (i)-(iii) for calibration events. This selection is sensitive to the momentum distribution of the primary electrons, and we obtain a relative systematic uncertainty of $11\%$ on the ratio of the yield $n_{\mathrm{MC}}/n_{\mathrm{data}}$.

In order to compare the number of remaining events, we normalize both samples to the same number of EOT. This is done by scaling the MC sample with the following factor:

\begin{equation}
    f = \frac{N_{\mathrm{EOT, data}}}{N_{\mathrm{EOT, MC}}\times b}
\end{equation}

where $b$ the bias factor on the electron-nuclear cross-section in MC, and $N_{\mathrm{EOT, data}}$ ($N_{\mathrm{EOT, MC}}$) is the number of EOT in data (MC).

\begin{figure}[ht!]
    \centering
    \includegraphics[width=0.475\textwidth]{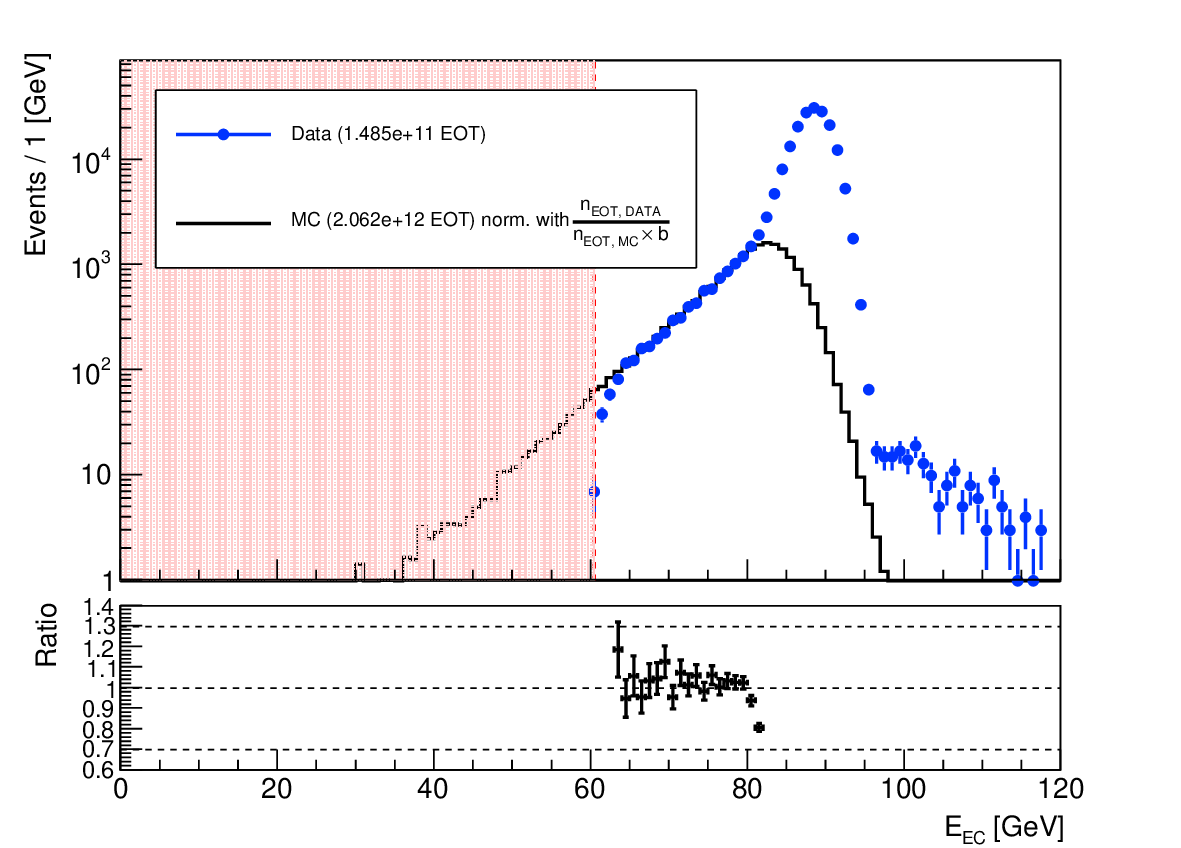}
    \caption{ECAL energy spectrum for events in the sideband $E_{\mathrm{HC}}<\SI{1.5}{GeV}$ plane before applying a cut on the total VHCAL energy for the blinded sub-period of the 2023 run and the MC sample, after the selection criteria (i)-(vi) are applied. The MC sample is normalized to the number of EOT in the sub-period with the factor $f$ as described in the text. The light-red shaded area corresponds to the signal region in the ECAL plane. The bottom plot displays the ratio $\mathrm{MC}/\mathrm{data}$ for each bin.}\label{fig:comparison_MC_data_ECAL_analysis} 
\end{figure}

First, we compare the events remaining after a subset of the invisible selection criteria (i)-(vi), as described in Section~\ref{sec:NA64}. We skip the cuts on the multiplicity of hits in ST3 and ST4 (vii) and the hadron calorimeter response (viii), as the purpose of these cuts is to remove EUM events. As demonstrated in Section \ref{subsec:Data analysis}, events in the sideband $E_{\mathrm{HC}}<\SI{1.5}{GeV}$ with energies below $E_{\mathrm{EC}} < \SI{80}{GeV}$ are effectively removed by the VHCAL cut and correspond to $e^{-}$-nuclear interactions before the ECAL. As shown in Fig.~\ref{fig:comparison_MC_data_ECAL_analysis}, this is confirmed by the MC sample, as the region populated by these events in the ECAL spectra matches closely with what is observed in data. 

In the ECAL energy spectrum, the missing events below $E_{\mathrm{EC}}\lesssim\SI{62}{GeV}$ are due to the blinding applied to the data sample. This blinding process consists of masking the signal region in order to ensure that our cuts remain unbiased with respect to the measurement of these rare processes. The large peak in data at $E_{\mathrm{EC}}\approx\SI{90}{GeV}$ originates from the low-energy tail of the $\SI{100}{GeV}$ primary electrons that pass the production trigger $\mathcal{S}_{phys}$. Considering the systematic uncertainty associated with the description of the beam, the ratio of the yield in MC with respect to the 2023 run data for the interval $E_{\mathrm{EC}} \in [\SI{60}{GeV}, \SI{80}{GeV}]$ is $n_{\mathrm{MC}}/n_{\mathrm{data}}=\SI[parse-numbers=false]{1.02 \pm 0.01 (stat) \pm 0.11 (sys)}{}$, after normalizing both samples to the same number of EOT. The $10\%$ difference can be explained by missing material due to the idealized definition that leads to a mismatch in the material budget in the MC simulation and is well within the systematic uncertainty associated with the beam description. 

\begin{figure}[ht!]
    \centering
    \includegraphics[width=0.44\textwidth]{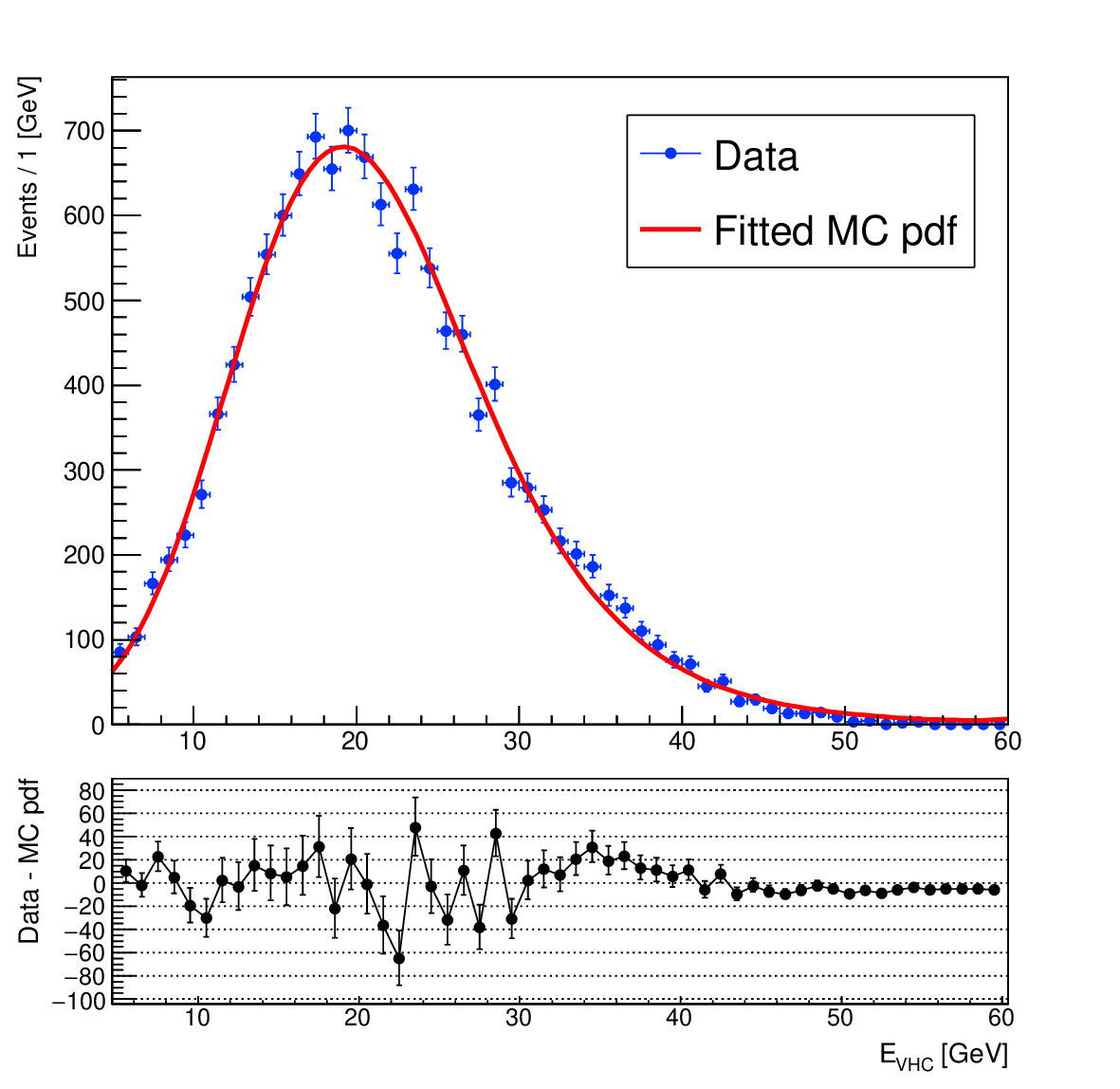}
    \caption{VHCAL spectrum for all events before applying a cut on the total VHCAL energy for the blinded sub-period of the 2023 run and the MC sample, after all other selection criteria are applied. The probability density function (pdf) of the MC sample is fitted to the data after the convolution with a Gaussian distribution using \texttt{RooFit}. The bottom plot displays the difference $\mathrm{data}-\mathrm{MC}$ for each bin.}\label{fig:comparison_MC_data_VHCAL_analysis} 
\end{figure}

We can also compare the observed VHCAL energy spectra obtained without the cut on $E_{\mathrm{VHC}}$. This is illustrated in Fig.~\ref{fig:comparison_MC_data_VHCAL_analysis}, where we find that a good fit is obtained by convolution of the MC probability density function with a normal distribution. This is consistent with a slight mismatch in the relative position of the electron-nuclear interaction point and the VHCAL position. The resulting difference in the VHCAL acceptance leads to a reduced energy in MC by less than $\SI{3}{GeV}$.

\begin{figure*}[!ht]
\begin{center}
\begin{subfigure}{.33\textwidth}
    \centering
\includegraphics[trim={0cm 0cm 0cm 1.2cm},clip,width=0.95\textwidth]{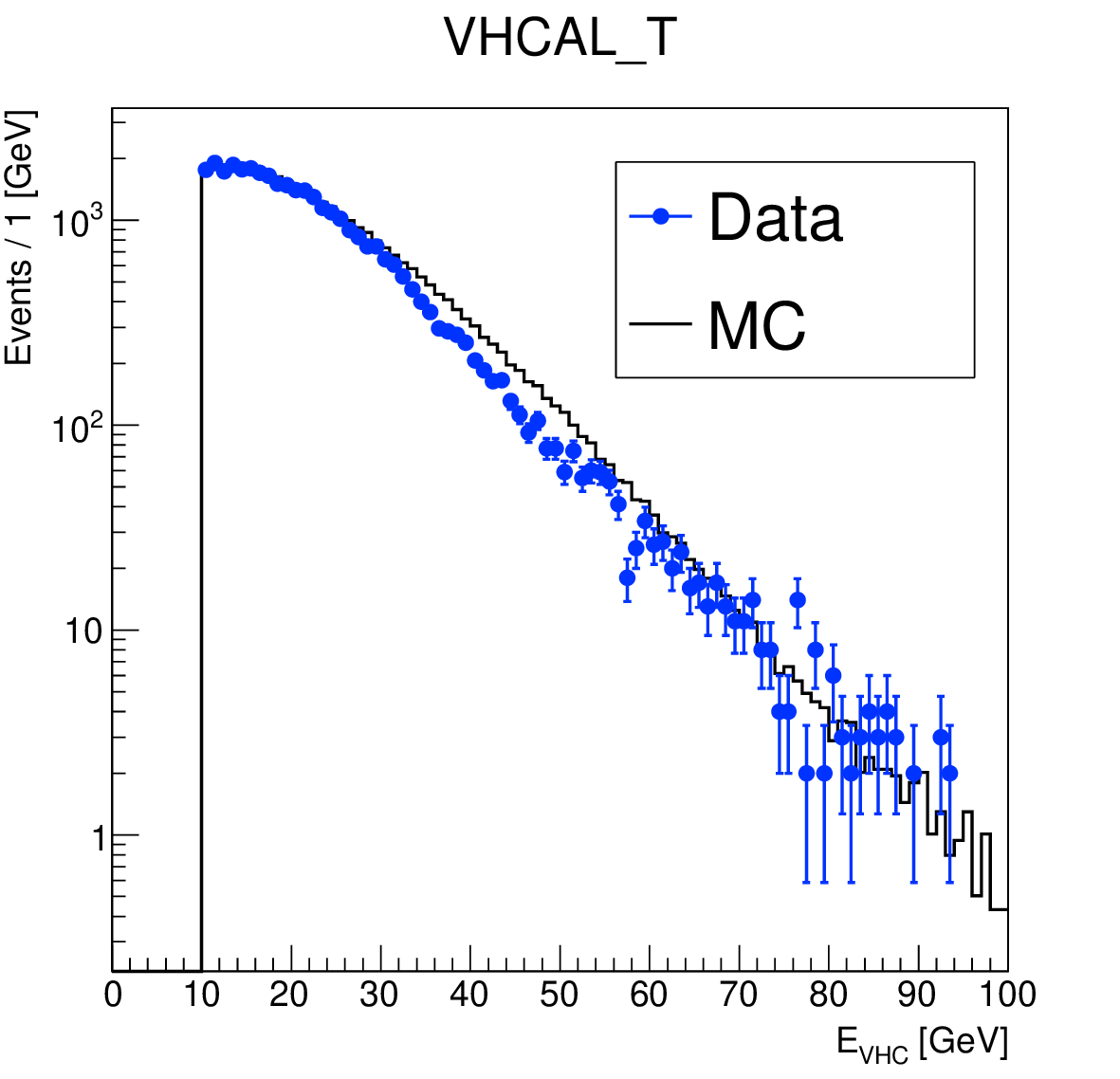}
\end{subfigure}%
\begin{subfigure}{.33\textwidth}
    \centering
    \includegraphics[trim={0cm 0cm 0cm 1.2cm},clip,width=0.95\textwidth]{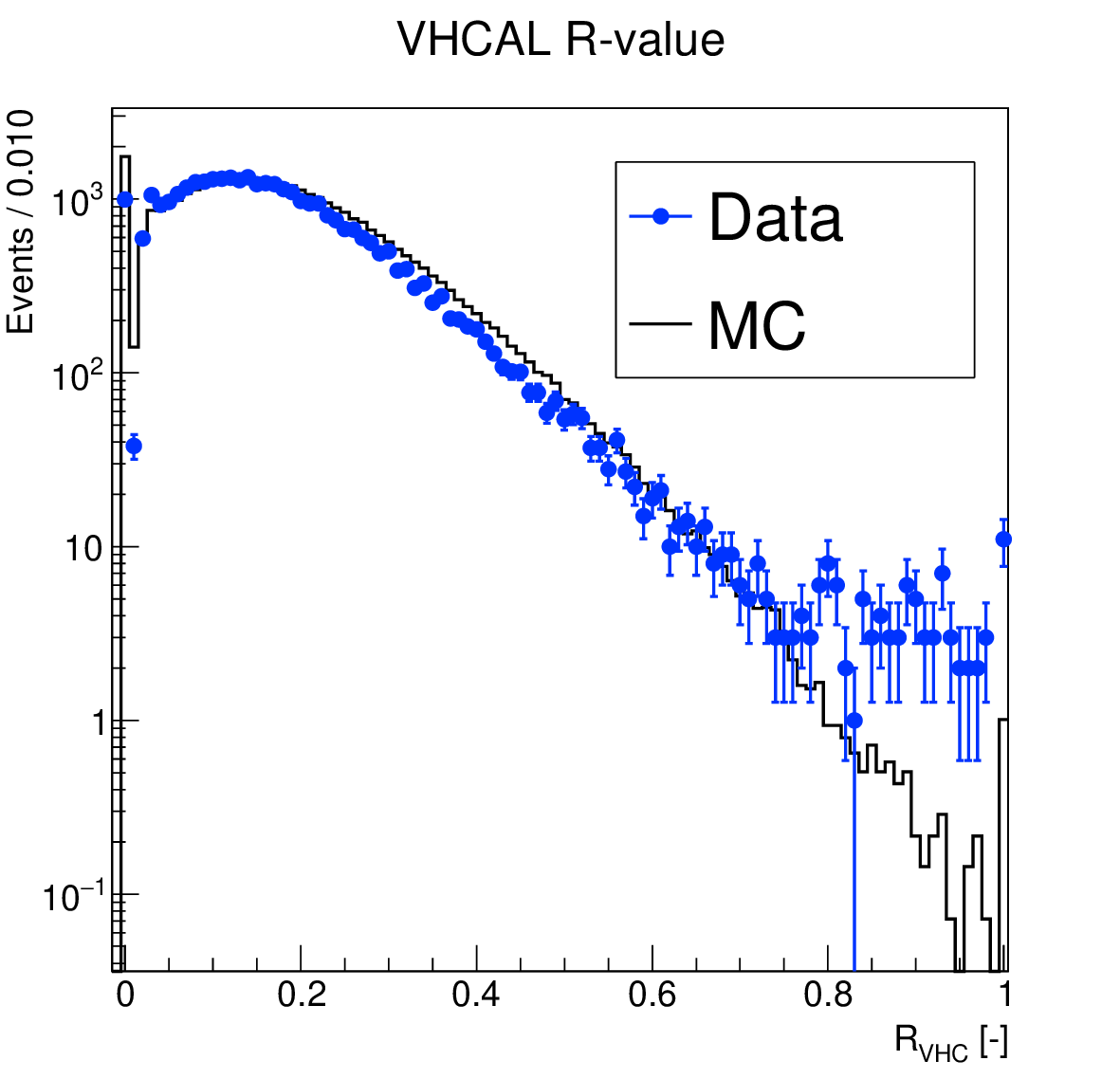}
\end{subfigure}%
\begin{subfigure}{.33\textwidth}
    \centering
    \includegraphics[trim={0cm 0cm 0cm 1.2cm},clip,width=0.95\textwidth]{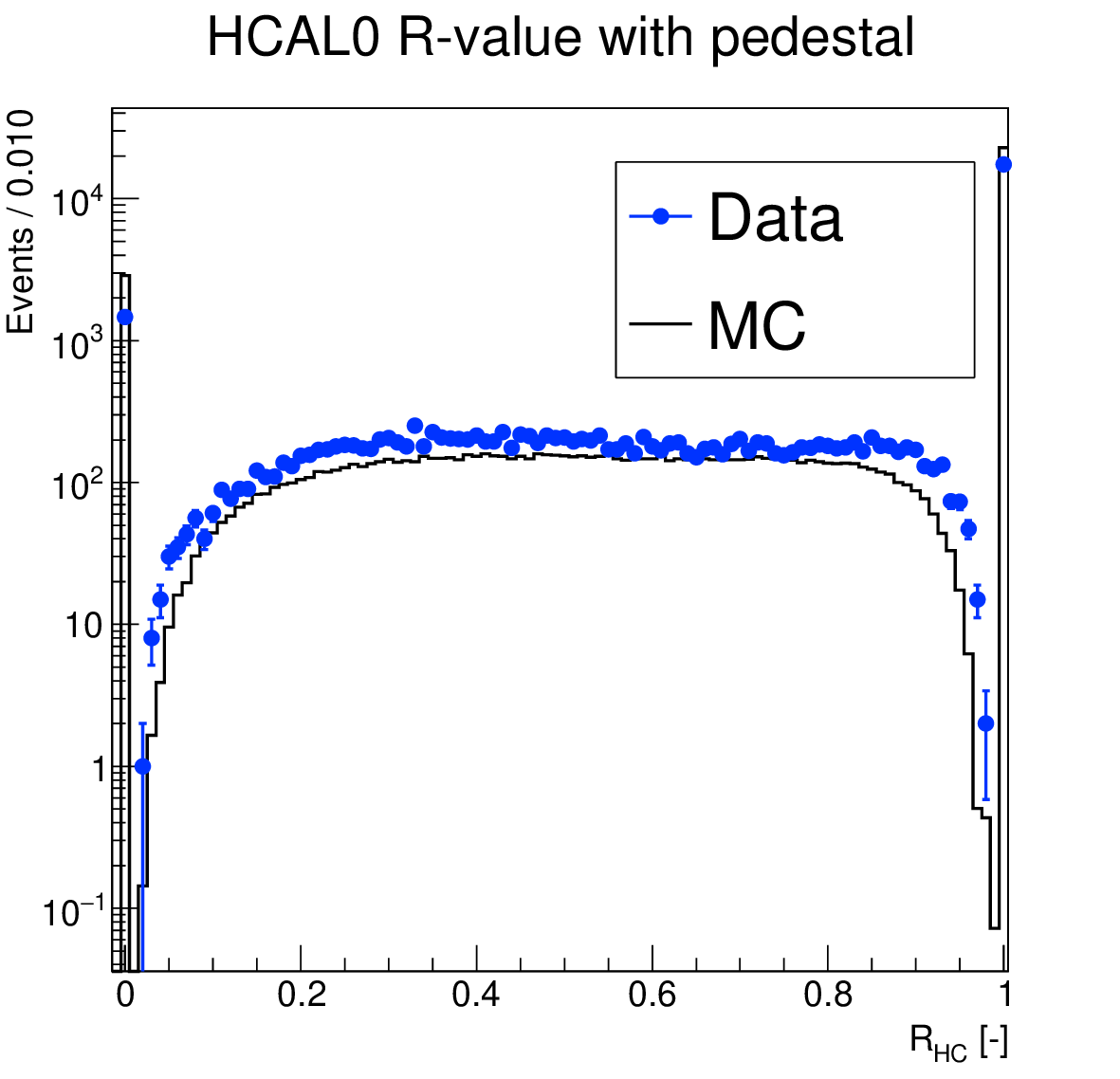}
\end{subfigure}
\end{center}
    \caption{Comparison of the VHCAL energy spectrum (left), $R_{\mathrm{VHC}}$ (center) and $R_{\mathrm{HC}}$ (right) for events with energy deposition in VHCAL, VETO and the first HCAL module. The MC sample is normalized to the number of EOT in the sub-period with the factor $f$ as described in the text.}
    \label{fig:comparison_MC_data_leak} 
\end{figure*}

To establish how well the response of the VHCAL is reproduced, we exclusively compare EUM events with secondary hadrons that traverse the VHCAL, VETO and first HCAL modules. As sketched in Fig.~\ref{fig:VHCAL_coverage}, in this case the scattered beam electron impinges on the ECAL and one or more secondary particles are detected in the VETO and in both hadronic calorimeters (VHCAL and HCAL). In this way, we aim to select the subset of events with a leading charged hadron, mostly $\pi^{-}$ or $\pi^{+}$, from $e^{-}$-nuclear production in the vacuum window, $S_{2}$, $V_{1}$ and MM3. To achieve this, we consider a reduced selection with SRD (i), ECAL (iv) and HCAL$_{4}$ (v) cuts, but without track information. Instead, we require significant activity in the prototype VHCAL $E_{\mathrm{VHC}}\gtrsim\SI{10}{GeV}$ accompanied by the detection of a MIP in VETO and an energy deposition above $\SI{1}{GeV}$ in HCAL$_{1}$.

As shown in Fig.~\ref{fig:comparison_MC_data_leak}, the simulation reproduces the response of the calorimeters for these events, both in terms of energy and shower shape. In contrast to the previous selection, no corrections to the VHCAL energy are needed in MC, as the energy requirement in both VHCAL and HCAL imposes a stricter condition on the angular distribution of the secondary hadrons. Therefore, any effect stemming from discrepancies in the VHCAL acceptance is reduced. Moreover, the total efficiency after this selection matches well in both MC and data after all cuts, and we obtain $n_{\mathrm{MC}}/n_{\mathrm{data}}=\SI[parse-numbers=false]{1.12 \pm 0.02 (stat) \pm 0.11 (sys)}{}$.

In conclusion, simulations of $e^{-}$-nuclear interactions using Geant4 as the event generator reproduce the observed response from the production trigger events in 2023. Therefore, simulations with biased electron-nuclear interactions can be considered to provide a good estimate within an $11\%$ uncertainty on the expected contribution to the background from large-angle hadrons for different configurations and a larger number of EOT. This step is crucial to develop an improved setup, capable of dealing with the background expected at $10^{13}$~EOT.

\section{Results}\label{sec:Results}

\begin{figure}[ht!]
\begin{center}
\begin{subfigure}{.155\textwidth}
    \centering
    \includegraphics[width=0.95\textwidth]{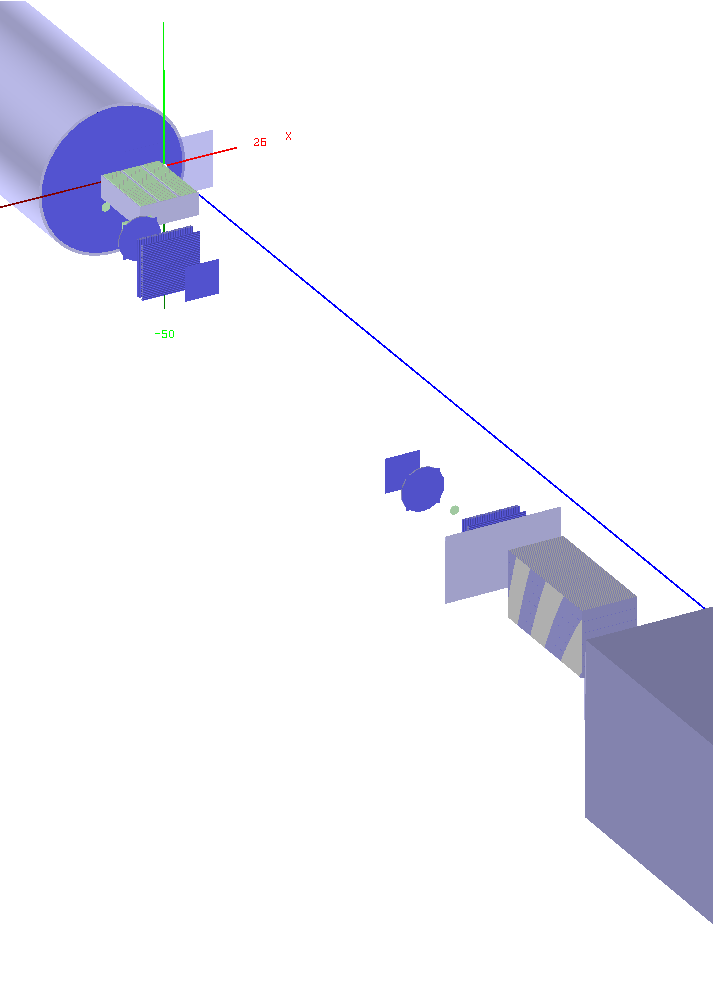}
\end{subfigure}
\begin{subfigure}{.155\textwidth}
    \centering
    \includegraphics[width=0.95\textwidth]{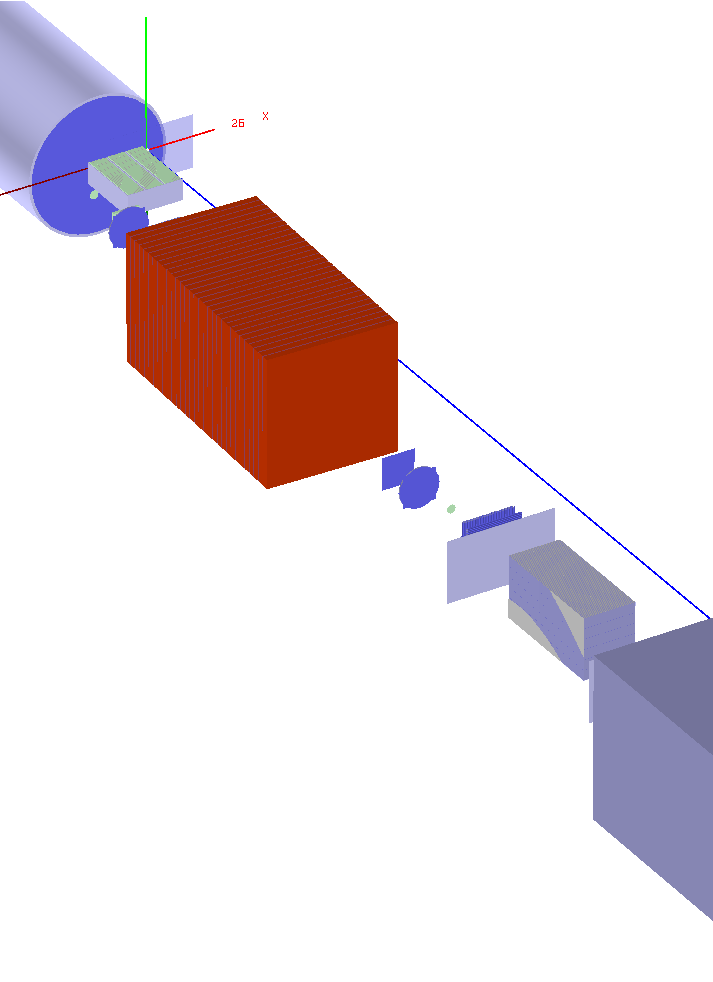}
\end{subfigure}%
\begin{subfigure}{.155\textwidth}
    \centering
    \includegraphics[width=0.95\textwidth]{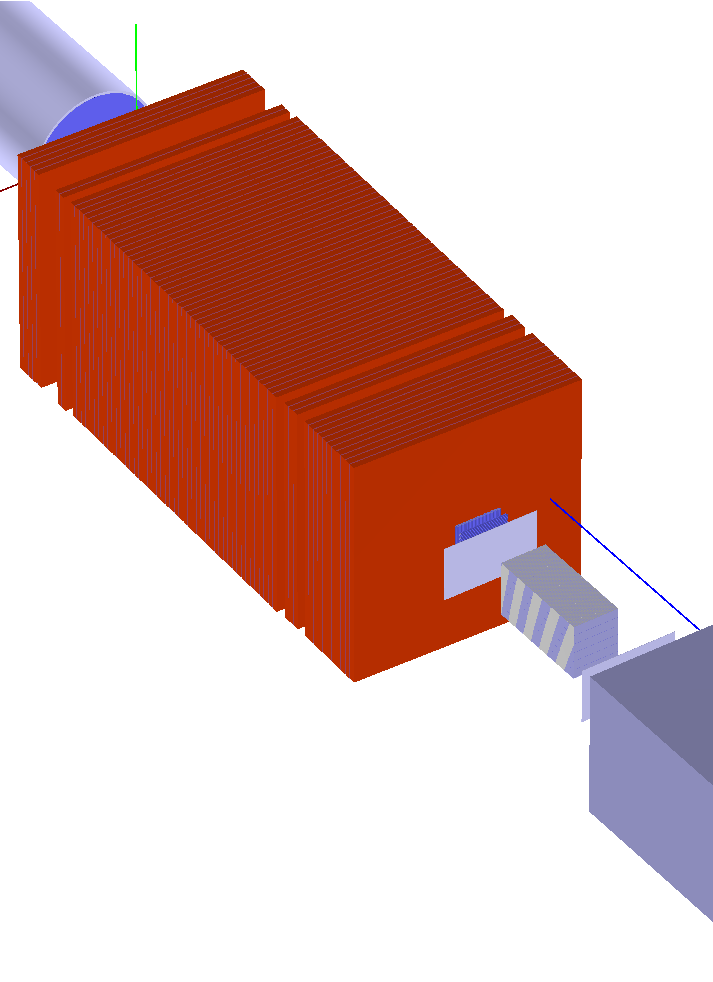}
\end{subfigure}
\end{center}
\caption{A rendering of the simulated geometry without (left), with the current prototype tested in the 2023 run (middle) and with a full-scale (right) VHCAL. Drawn are the axes at the end of the vacuum vessel, with the z axis (blue) corresponding to the unbent beam direction.}
\label{fig:setup_MC} 
\end{figure}

Having motivated the usage of the NA64 MC framework to study the EUM background and improve the current electron mode setup, we present the results for the three different configurations: one with the VHCAL removed from the simulated geometry, another with the realistic prototype VHCAL as it was placed in 2023 and one with a full-scale VHCAL that covers almost all the downstream region. In this case, the dimensions of the larger VHCAL are approximately $1000\times1000\times\SI{2000}{mm^3}$, doubling the length of each side and including layers between the trackers to cover the entire downstream region. This configuration could be achieved by having the trackers inside the calorimeter or placing several shorter modules in between the trackers, therefore ensuring that the quality of the track reconstruction is preserved. All these configurations are shown in Fig.~\ref{fig:setup_MC}. 

Before discussing the differences between these three setups, we remark on some common characteristics of the hadronic background in the simulated samples. For all three samples, we obtain compatible trigger and hadronic electroproduction rates. In particular, for $e^{-}$-nuclear and $\gamma$-nuclear interactions with a transferred energy $E_{\mathrm{transfer}} \geq \SI{10}{GeV}$, we observe a relative fraction $n_{en}/n_{\mathcal{S}_{trig}} \approx 0.73\%$ including the bias. Even though the bias is applied on all volumes upstream of the ECAL, more than $\approx99.95\%$ of the events saved have an interaction vertex in the downstream region of the detector. That is, the interaction takes place between the end of the vacuum vessel and the beginning of the ECAL. Moreover, $56\%$ of these interactions happen after the end of the VHCAL. This is because events for which hadronic production takes place in or very close to $S_{3}$ are more likely to have a scattered electron that passes the trigger condition $\mathcal{S}_{trig}$.

Conversely, the probability for an event with secondary hadrons from $e^{-}$-nuclear interactions upstream of the magnet spectrometer to pass the trigger condition $\mathcal{S}_{trig}$ is less than $0.05\%$. The background rate for events with a production vertex before the bending magnets remains mostly constant regardless of the VHCAL configuration, as the coverage in the downstream region is not enough to detect this type of event. After closely investigating these events, we observe that they are mostly characterized by a larger spread in the reconstructed entrance angle and higher multiplicity of hits in ST2, suggesting possible ways to identify these interactions.

\begin{figure*}[!ht]
\begin{center}
\includegraphics[trim={0 0 0 0.9cm},clip,width=0.8\textwidth]{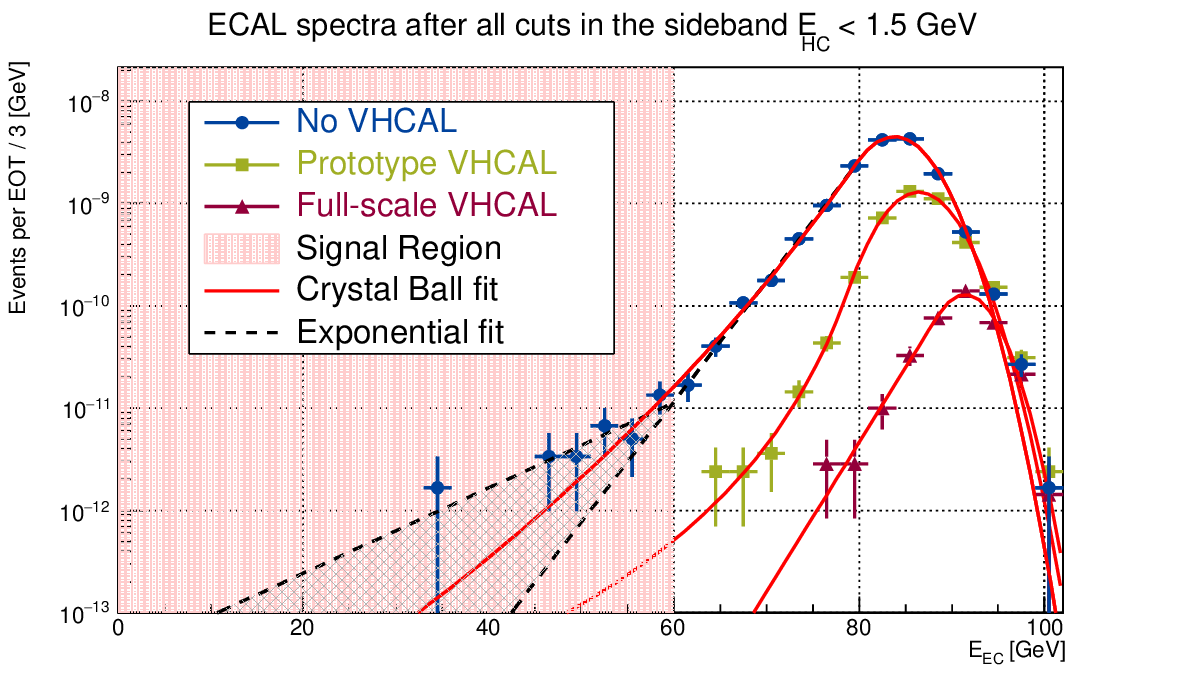}
    \caption{ECAL spectra of the three simulated configurations for the remaining events in the sideband $E_{\mathrm{HC}}<\SI{1.5}{GeV}$ after applying all invisible selection criteria (i)-(viii). All MC samples were normalized to display the fraction of events per EOT. The light-red shaded area corresponds to the signal region in the ECAL plane. The fit of the remaining distribution of events using the Crystal Ball function are shown with solid, red lines. The dashed, black lines illustrate two exponential fits for the MC sample used to determine the systematic uncertainty from the choice of fitting function and range.}
    \label{fig:comparison_MC_extrapolation} 
\end{center}
\end{figure*}

Returning to the comparison of the hermeticity in these configurations, Fig.~\ref{fig:comparison_MC_extrapolation} illustrates the resulting $E_{\mathrm{EC}}$ distribution for the three MC samples. The comparison is presented after applying the production trigger selection $\mathcal{S}_{phys}$ and all invisible selection criteria (i)-(viii), except for cut on the e-m shower in the ECAL. We fit the remaining events with a Crystal Ball function and determine from its integral over the signal region the expected background contribution for the different setups. In addition to this, we estimate the systematic uncertainty from the choice of the fitting function and range by fitting an exponential function to the sample without VHCAL, as shown by the dashed, black lines in the plot. For the configurations involving the prototype and the full-scale VHCAL, we conservatively assume the same relative systematic uncertainty, as there are insufficient data points left to perform a similar procedure.

As expected, the setup without a VHCAL lacks the necessary coverage to ensure a low background. On the other hand, the most important finding of this study is that there is substantial potential for improvement, which is made obvious by comparing the extent of background suppression in the case of a full-scale VHCAL. This full-scale VHCAL allows reducing the expected EUM background by at least an additional order of magnitude. 

We summarize the estimates for the background contribution for these different setup configurations in Table~\ref{tab:summary}. As demonstrated in Section~\ref{subsec:MC validation}, the MC estimates provide a reliable description of EUM events and allow us to provide an estimate for the background per EOT. The resulting estimates for the prototype VHCAL case translate to a conservative contribution of $n_{b} \approx 0.4$ for $4.38\times10^{11}$~EOT in 2023, which is expected to be lower given the missing e-m shower cut. Although the current prototype VHCAL represents a significant improvement, achieving a background-free setup at the level of $10^{13}$~EOT - particularly concerning escaping hadrons from EUM - will require an upgraded VHCAL module. Overall, these findings are promising and validate the current approach of NA64.

\begin{table}[h]
\small
    \renewcommand{\arraystretch}{1.2}
    \centering
    \begin{tabular}{l|l}
    \toprule
    \toprule
    Configuration & $n_{b}$ per EOT\\
    \hline
      No VHCAL    &  $\SI[parse-numbers=false]{(3.0 \pm 2.0 (stat) \pm 1.0 (sys))\times10^{-11}}{}$ \\
      Prototype VHCAL    & $\SI[parse-numbers=false]{(1.0 \pm 6.0 (stat) \pm 0.3 (sys))\times10^{-12}}{}$ \\
      Full-scale VHCAL & $\SI[parse-numbers=false]{(1.0 \pm 2.0 (stat) \pm 0.3 (sys))\times10^{-14}}{}$ \\
    \bottomrule
    \bottomrule
    \end{tabular}
    \caption{Summary of the background estimates per EOT from extrapolation of the fitted distribution in all three MC configurations after applying the invisible selection criteria (i)-(viii). See Section~\ref{sec:NA64} for a description of the cuts.}
    \label{tab:summary}
\end{table}

\section{Conclusions}\label{sec:Conclusions}

In this work, we assess the impact of the prototype VHCAL as a measure to reduce the EUM background. Preliminary data-driven estimates indicate that the introduction of the VHCAL in the setup has resulted in at least an order-of-magnitude in background suppression compared to 2022. Furthermore, we find that MC simulations performed with Geant4 accurately reproduce the detector response to hadrons generated in EUM events, showing excellent agreement between the remaining events in the simulation and the data from the 2023 run. Specifically, the energy deposition and the transverse distribution of the hadronic showers are well modeled for electron-nuclear interactions with large energy transfer. In this way, we propose a MC-based approach to estimate the suppression of this background in present and future experimental configurations. All in all, these simulations suggest that with an optimized, full-scale VHCAL, a background suppression below $1$ event per $10^{13}$~EOT could be achievable. 

These findings provide critical insights into the types of particles escaping the detection in the current invisible setup and offer valuable guidance for designing the next generation of experimental upgrades. Although the suppression of background from other SM interactions is beyond the scope of this work, complementary methods to achieve this need to be investigated as well. We reiterate that increased hermeticity has a dual benefit, as it allows reducing the expected background and increasing the signal region. This provides a further enhancement in the search for DM, accelerating the progress of NA64.

As presented in Section~\ref{sec:Introduction}, these results are crucial in the long-term plans of the NA64 experiment, both in its electron \cite{Crivelli:2907892} and positron beam configurations \cite{Andreev:2025positron}, and confirm our understanding of this background. Further studies are ongoing, seeking to complete the design of a full-scale VHCAL that can provide the required coverage to remain background-free in the runs after LS3.

\section*{Declaration of competing interest}
The authors declare that they have no known competing financial interests or personal relationships that could have appeared to influence the work reported in this paper.

\section*{Data availability}
Data will be made available on request.

\section*{Acknowledgments}
We gratefully acknowledge the support of the CERN management and staff and the technical staff of the participating institutions for their vital contributions. This work was supported by the HISKP, University of Bonn (Germany), ETH Zurich Grant No. 22-2 ETH-031, and SNSF Grant No. 186181, No. 186158, No. 197346, No. 216602, No. 219485 (Switzerland), and FONDECYT (Chile) under Grant No. 1240066 and Grant No. 3230806, and ANID - Millenium Science Initiative Program - ICN2019 044 (Chile), and RyC-030551-I and PID2021-123955NA-100 funded by MCIN/AEI/ 10.13039/501100011033/FEDER, UE (Spain), and COST Action COSMIC WISPers CA21106, supported by COST (European Cooperation in Science and Technology). This result is part of a project that has received funding from the European Research Council (ERC) under the European Union's Horizon 2020 research and innovation programme, Grant agreement No. 947715 (POKER). This work is partially supported by ICSC – Centro Nazionale di Ricerca in High Performance Computing, Big Data and Quantum Computing, funded by European Union – NextGenerationEU.

\bibliography{bibliographyNA64_inspiresFormat,bibliographyNA64exp_inspiresFormat,bibliographyOther_inspiresFormat}

\end{document}